# Finnish 5th and 6th graders' misconceptions about Artificial Intelligence

Dr. Pekka Mertala (corresponding author)
- Department of Teacher Education, University of Jyväskylä, Jyväskylä, Finland
- Alvar Aallon katu 9, 40014, Jyväskylä, Finland
- pekka.o.mertala@jyu.fi

Dr. Janne Fagerlund
- Finnish Institute for Educational Research, University of Jyväskylä, Jyväskylä, Finland
- Alvar Aallon katu 9, 40014, Jyväskylä, Finland
- janne.fagerlund@jyu.fi

## Abstract

Research on children's initial conceptions of AI is in an emerging state, which, from a constructivist viewpoint, challenges the development of pedagogically sound AI-literacy curricula, methods, and materials. To contribute to resolving this need in the present paper, qualitative survey data from 195 children were analyzed abductively to answer the following three research questions: What kind of misconceptions do Finnish 5th and 6th graders' have about the essence AI?; 2) How do these misconceptions relate to common misconception types?; and 3) How profound are these misconceptions? As a result, three misconception categories were identified: 1) Non-technological AI, in which AI was conceptualized as peoples' cognitive processes (factual misconception); 2) Anthropomorphic AI, in which AI was conceptualized as a human-like entity (vernacular, non-scientific, and conceptual misconception); and 3) AI as a machine with a pre-installed intelligence or knowledge (factual misconception). Majority of the children evaluated their AI-knowledge low, which implies that the misconceptions are more superficial than profound. The findings suggest that context-specific linguistic features can contribute to students' AI misconceptions. Implications for future research and AI literacy education are discussed.

## Research highlights
- Students possess different kind of AI-misconceptions
- These misconceptions vary with regards to their fundamentality and type.
- Linguistic features can contribute to students' AI- misconceptions.
- Students' AI-misconceptions are more superficial than profound

## 1. Introduction

Artificial intelligence (AI) is a highly ubiquitous and pervasive digital technology. Different kinds of AI applications are used in (almost) every aspect of human life including health (medical diagnoses), education (learning analytics), household tasks (robot vacuums), information retrieval (web searches), and recreation (digital games) to provide only a few examples. As a result, it has been argued that agentic subjectivity in an age of AI requires AI-literacy (e.g., Lee et al., 2021), one dimension of which is to understand what AI is (Long & Magerko, 2020; Nq et al., 2021). Recently, AI-literacy has been introduced in educational curricula and white papers in various countries (e.g., Su et al., 2022) and scholars have begun to create methods and materials on how children could be taught about AI (e.g., Druga & Ko, 2021; Druga et al., 2022; Iregens et al., 2022; Kim et al., 2023; Lee et al., 2021; Shamir & Levin, 2022; Su et al., 2022; Vartiainen et al., 2021). That said, without knowledge about children's initial conceptions, the design of relevant curricula has less than solid footing.

From a constructivist viewpoint, the benefit of studying students' initial conceptions is that it tells us about the possible misconceptions they may possess. If students have a misconception prior to learning a subject, this may prevent them from learning the new subject properly, thereby leading to new misconceptions (Biber et al., 2013). Research on children's initial conceptions of abstract digital technologies like computers (Rucker & Pinkwart, 2016) the Internet (Babari et al., 2023; Eskelä-Haapanen & Kiili, 2019), search engines (Kodama, 2016) the Internet of Things (Mertala, 2020), digital data (Agesilaou, & Kyza, 2022; Pangrazio & Selwyn, 2018), and programming / coding (Mertala, 2019) suggest that informal encounters provide children only a limited understanding of what these technologies actually are. AI as an "opaque technology" (Long & Magerko, 2020) and a "fuzzy concept" (Kaplan & Haenlein, 2019) should not be an exception. Thus, acknowledging students' preconceptions is seen as a fundamental prerequisite for pedagogically sound AI-literacy education (Kim et al., 2021; Long & Magerko, 2020).

Research on children's initial conceptions of AI is in emerging state and has been touched upon mainly with small samples (N=8–17; Kreinsen & Schultz, 2021; Ottenbreit-Leftwich et al., 2021; Ottenbreit-Leftwich et al., 2022; Solyst et al., 2023; Vandenbergen & Mott, 2023; cf. Mertala et al., 2022) and sometimes with no data excerpts that would shed light on the rationales underlying children's thinking (Kreinsen & Schultz, 2021; Ottenbreit-Leftwich et al., 2021). Additionally, some studies were conducted within AI-themed summer camps and workshops, and the participating children were reported to have prior experience with programming and were interested in computer science (Kim et al., 2023; Solyst et al., 2023), which –alongside with a small sample— limits the transferability of the findings.

Furthermore, the instructions in some studies have contained elements we refer to as conceptual priming. Students in Vartiainen et al. (2021) study were given an introductory task to report "what they knew or thought they knew about artificial intelligence". The exact instruction was to "draw and/or write [...] thoughts and ideas about *how one could teach a computer*" (italics original; Vartiainen et al., 2021). Asking children to explain what AI is most likely provides different answers than asking them to describe how computers can be taught. The latter instruction contains a rather explicit cue that computers can be taught, which arguably leads to more unified responses than the use of more open-ended questions (Vosniadou & Brewer, 1992; see also Selwyn et al., 2022). Last, the existing research has not touched upon the question of how profound or superficial children's AI misconceptions are.

In sum, there is a need to study children's misconceptions about AI with 1) larger samples, 2) instructions that avoid conceptual priming, and 3) approaches that shed light on the profoundness of misconceptions. This paper contributes to resolving these needs by exploring 195 Finnish 5th and 6th graders' (12–13-year-olds') misconceptions about AI via an open-ended qualitative survey. The questions guiding the research process are:

1. What kind of misconceptions do Finnish 5th and 6th graders' have about the essence of AI?
2. How do these misconceptions relate to common misconception types?
3. How profound are these misconceptions?

## 2. Background: Misconceptions (about AI)

A prerequisite for researching misconceptions about AI is to define what AI and (mis)conceptions are. Here, we draw on Kurzweil's (1990) definition of AI as the art of creating machines that perform functions that require intelligence when performed by people. Although the definition dates back to the 1990's, it provides a solid starting point for identifying AI-related misconceptions. First, it suggests that AI is not intelligent per se. Instead, it is capable of successfully conducting tasks that are considered requiring human intelligence (e.g., image recognition and sorting, composing sensible text). Second, by highlighting the role of machines in mimicking human intelligence, the definition emphasizes that AI is bound to digital technology: "AI cannot be done with a pencil and piece of paper, hence, a computer is always required", as Emmert-Streib et al. (2020, n.p.) neatly put.

Conceptions, in turn, refer to ideas that people use to understand the world around them (Marton, 1981; Thompson & Logue, 2006). Put differently, conceptions are explanations and hypotheses of what things are: how they come to be, how they operate, and so on. Correct explanations are often referred to as scientific conceptions (Vygotsky, 1987), which provide a factual science-based explanation for how and why things work and what they are (Edwards et al., 2018). Misconceptions, then, can be described as ideas that provide an incorrect understanding of the world and its phenomena (e.g., Bahar, 2003; Smith et al., 1994).

The relationship between a correct conception and a misconception is not binary. Instead, the relationship can be thought of as a sliding scale. Many real-world phenomena are highly complex and thus extremely difficult to fully understand and describe even by professionals. Take the concept of "natural selection", for example: research suggests that natural selection is poorly understood not only by young students and members of the public but even among those who have had postsecondary instruction in biology (Gregory, 2009).

That being said, the conceptions of those who have studied biology in post-secondary level are most likely more accurate than those who have not and, thus, locate closer to the "correct" definition. This stands for AI as well: In a recent Australian survey, respondents with background in computer science were (self-reportedly) more cognizant about AI than others (Selwyn et al., 2022). However, defining what AI is, can be confusing even for experts as the term (and AI itself) has evolved over the course of time (Long & Magerko, 2020). Furthermore, misconceptions also vary with regards to degree of certitude (Usó-Doménech & Nescolarde-Selva, 2016): some misconceptions are superficial while others are more deeply held (i.e., preconceived notions, Davies, 1997).

While different sources use (partly) different terms and categories for describing the variety of misconceptions (e.g., Harlen & Qualter, 2018; Davis, 1997), they tend to share a common (often Vygotskian, [1987]) core that misconceptions are typically formed through everyday experiences. These experiences are often sense-based and, due to the lack of access to sensible alternative viewpoints, they provide only a partial evidence of the phenomenon (Harlen & Qualter, 2018).

Let us put statement that "AI methods work similar to the brain" —which according to Emmet-Streib et al., (2020, n.p) is a common belief— under a closer inspection. Such a view van be conceptualized as a false analogy, in which the similarity in one respect of two concepts, objects, or events is taken as sufficient to establish that they are similar in another respect in which they actually are dissimilar. Take ChatGPT and other AI-based chatbots, for example. They are similar people in a sense that both can produce sensible text. However, since ChatGPT creates new texts based on probabilities —it "guesses" which word is most likely to come next— it functions rather differently than human brain.

Equating AI with human brain can be further conceptualized as an anthropomorphic misconception, which refers to the attribution of distinctively human-like feelings, mental states, or behavioral characteristics to inanimate objects (Airenti, 2015; Epley et al., 2007). This is a common finding in research on children's (mis)conceptions of AI (Kim et al., 2023; Kreinsen & Schultz, 2021; Mertala et al., 2022) and digital technology in general (Rucker & Pinkwart, 2016; Robertson et al., 2018).

Anthropomorphic misconception can refer to different types of misconceptions outlined by Davis (1997). Anthropomorphic misconception of AI can be a *vernacular misconception*, that is, a language confusion where people mistake everyday speech lexemes for scientific terms (Keeley, 2012). More precisely, the term "intelligence", traditionally, has referred to human cognition and behavior but has a different meaning in the field of AI research and development than in colloquial language or other disciplines (Legg & Hutter, 2007).

Anthropomorphic misconception can also be a *non-scientific misconception* (Davis, 1997) that results from information retrieved from non-scientific sources. Anthropomorphic and agentic AI is a common representation of AI in popular culture (Cave & Dihal, 2019) and news media (Slotte Dufva & Mertala, 2021), and the influence of these imaginaries have been identified from peoples' conceptions of AI (Cave et al., 2019; Selwyn et al., 2022).

On the other hand, anthropomorphic misconception could be a *conceptual misconception* (Davis, 1997) that arises from limited experience and narrow focus (Harlen & Qualter, 2018) on specific features of AI. Children commonly name personal voice assistants like Siri and Alexa as an example of an AI they are aware or have experience of (Kim et al., 2023; Kreinsen & Schultz, 2021; Mertala et al., 2022; Vandenberg & Mott, 2023). They most likely encounter numerous other AI solutions in their daily life (e.g., recommendation algorithms, smart phones camera) but are not aware of the presence of AI in these applications – a common finding in studies done with adults (Selwyn et al., 2022; Zhang & Dafoe, 2020).

Lastly, anthropomorphic misconception can also be a *factual misconception* (Davis, 1997) – a failed attempt to reason the essence of a phenomenon. The more complex and opaque the technology, the more likely children rely on psychological reasoning (like anthropomorphism) when they explain the technology's functional capabilities (Turkle, 2003).

## 3. The present study

### 3.1. Data and participants

The data for the present study regarding students' misconceptions about AI were collected from 195 Finnish 5th and 6th graders from 10 different school classes via a qualitative online survey in April and May 2021. An invitation letter to participate openly and voluntarily by having students respond to an online questionnaire was sent to 5th and 6th grade teachers via email through a network of municipal school information and communication technologies coordinators in a medium-sized municipality in Central Finland. Ten teachers expressed interest in participating. As a requital and incentive for the time and effort invested in responding, the classes were provided with open web-based instructional material about the technical and ethical aspects of AI after the students had completed the survey. The study followed the practices of ethical research (Finnish National Board of Research Integrity, 2019) and current legislation on information privacy and data protection (GDPR.EU, 2022).

Our research motive in investigating the students' AI (mis)conceptions was exploratory and interpretive, focusing on the qualitative variety of the (mis)conceptions rather than aiming to form conception profiles linked with specific background variables. Thus, no personal data or sensitive information was collected from the participants. However, the students were asked whether they had participated in an ICT-themed club either as a hobby or as an elective subject in school. 11.6% of the students reported participating in such clubs. The most common examples were coding and gaming clubs. AI was not mentioned as a substance in the clubs. Furthermore, the students were asked to evaluate their knowledge regarding AI by responding to a statement "I know this subject [AI] well" on a five-step scale (1 = completely disagree, 5 = completely agree).

The questionnaire data for the present study consisted of the students' responses to five open-ended questions, which inquired about the students' conceptions of AI and the students' (numerical) self-evaluation of AI-related knowledge. In the questionnaire instructions, we emphasized (and instructed the teachers who were conducting the questionnaires) that the questionnaire should be completed alone, that it was not an exam or a test, and that there were no right or wrong answers. To avoid the priming effect identified in previous research (see Selwyn et al., 2022), we avoided value-laden expressions and statements. The questions posed in the questionnaire and their (literature-informed) justifications are presented in Table 1.

Table 1. Open-ended questions and their justification

| Questionnaire item | Rationale |
| --- | --- |
| 1. Describe what you think artificial intelligence means. | Provides information about how students understand AI either as a technology or as a concept (Long & Magerko, 2020). We chose to use the term "means" instead of "is" as previous research suggests that children may find it difficult to answer what an abstract concept or technology strictly "is" (Wennås Brante & Wallden, 2023). Thus, we reasoned that "means" would be a more inclusive and open-ended stem. |
| 2. Describe where you think artificial intelligence is or what it is used for. | Provides information about students' (mis)conceptions about the practical applications of AI and the contexts it is used (Kreinsen & Schultz, 2021) |
| 3. Describe how you think artificial intelligence works. | Provides information about students' (mis)conceptions about the technological / mathematical principles behind AI (e.g. Emmet-Streib et al., 2020; Vartiainen et al., 2021). |
| 4. Describe why you think artificial intelligence is used. | Provides information about students' (mis)conceptions of the motives behind the use (and development) of AI (Emmert-Streib et al., 2020). |
| 5. Name any words, things, or objects that you think are related to artificial intelligence | Provides possibility for dimensions not covered above in addition to free-association, which does not require students' to formulate full sentences. |

### 3.2. Analysis

The data was analyzed using an abductive approach that combines deductive and inductive reasoning (Grönfors, 2011). The role of deduction is to offer theoretical threads, which are complemented and/or refined via interpretive inductive analysis (Mertala, 2020). As a result, inductive approach forms a "hermeneutic circle" of reading, reflective writing, and interpretation (see, Kafle, 2013). In our case, the main theoretical threads were: 1) Kurzweil's (1990) definition of AI (AI as the art of creating machines that perform functions that require intelligence when performed by people); 2) Davis' (1997) types of common misconceptions (preconceived notion, scientific misconception, conceptual misconception, vernacular misconception, factual misconception), and 3); the prevalence of anthropomorphic misconceptions (e.g., Kim et al., 2023; Kreinsen & Schultz, 2021).

In the first phase of the analysis, the data (raw student responses to the questions) were read holistically. Each individual student's responses to each question were read and coded as a whole, that is, with respect to responses to all the questions. Student 23, for instance, replied that "it works with reason" to the third question ("Describe how you think artificial intelligence works"). This answer in itself provides no clues whether the student is talking about AI as technology, algorithms, or as something else (see the section 4.1. Non-technological AI for further discussion). However, the student's reply to the first question ("Describe what you think artificial intelligence means") allowed us to place the answer in the context: the student wrote that "some kind of robots can be artificial intelligence" and specified that AI-enabled robots are a distinct "form of life". Based on this information, the excerpts were placed under the category of anthropomorphic AI because AI was equated with sentient living creatures (like humans). Student 23's response was also coded as a vernacular misconception, because the student makes no difference between AI and human intelligence (AI is "form of life" that is able to "reason"). As a result of the analysis of all responses, three misconception categories with elements from (partly) different misconception types were formed (see Figure 1).

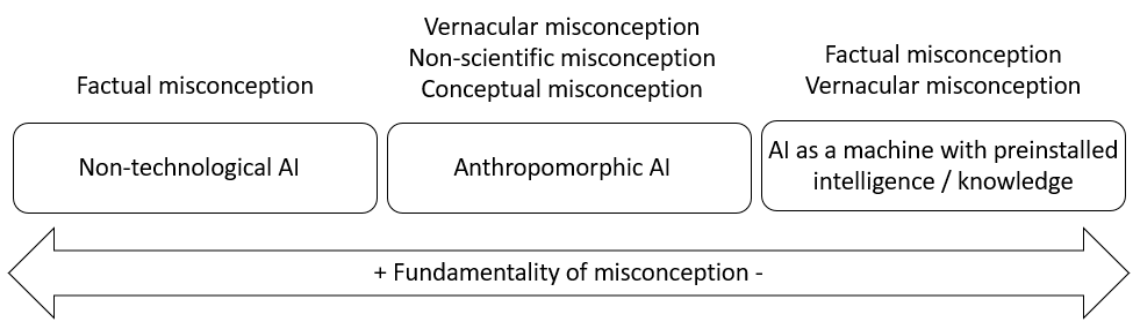

Figure 1: A synthesis of students' misconception types regarding AI

The second phase in our analysis was evaluating these categories on a scale of fundamentality from most fundamental to least fundamental by comparing them with Kurtzweil's (1990) definition of AI. Non-technological misconceptions were ranked as the most fundamental ones as they (more misinformedly) describe AI as human decision-making without any references to digital technologies. AI as a machine with pre-installed knowledge or intelligence, in turn, was ranked as the least fundamental one. In this category, AI was (less misinformedly) understood to be a technology, which can carry out processes that mimic intelligence, but the functional principles of AI were misunderstood. Note that fundamental here refers to the (metaphorical) distance between the misconception and a correct understanding of the concept, and it says nothing about the intensity or stability of the (mis)conception.

In the third phase we included the students' self-evaluations regarding their AI-related knowledge to gain insight about the profoundness of their misconceptions. Students who had evaluated their knowledge high (4 or 5 out of 5) were coded to possess a high level of certitude (Usó-Doménech & Nescolarde-Selva, 2016). Such students are marked with an asterisk (*) in the Findings and discussion section. The section also presents more data excerpts and the rationales behind the analysis process to improve the transparency of the analysis process. Some narrative soothing (Polkinghorne, 1995), such as correcting misspelled words (e.g., Aple → Apple), is done to improve the narrative flow of the data extracts.

## 4. Findings and discussion

The main findings of the study are presented in three subsections starting with the most fundamental misconceptions (see Figure 1). Discussion with previous relevant literature is integrated within the presentation of the results. The relationship between the misconception category (non-technological, anthropomorphic, pre-installed) and type (factual, non-scientific, vernacular, and conceptual) is outlined within the subsections. Table 2 summarizes the distribution of the different misconception categories, with anthropomorphic AI being the most common.

Table 3. Distribution of different misconception categories

|  | Non-technological AI | Anthropomorphic AI | AI as a machine with preinstalled intelligence / knowledge |
|---|---|---|---|
| **n** | 10 | 35 | 12 |
| **%** | 5,1% | 18% | 6,2% |

Regarding the level of certitude of the misconceptions, as outlined in Table 3, only 18.4% of the students thought that they had good knowledge about AI (self-evaluated knowledge response: 4 or higher). 38.5% of the participants, in turn, evaluated their knowledge low (self-evaluated knowledge response: 2 or lower). Additionally, 32 students (16.4%) responded "I don't know" / "I can't explain" or equivalent to both questionnaire items 1 (describe what you think artificial intelligence means) and 3 (describe how you think artificial intelligence works), which were about the essence of AI.

Table 3: Students' self-evaluated knowledge regarding AI on a five-step scale

|  | 1 | 2 | 3 | 4 | 5 | Missing data* | Total |
|---|---|---|---|---|---|---|---|
| **n** | 37 | 38 | 76 | 30 | 6 | 8 | 195 |
| **%** | 19% | 19.5% | 39% | 15.3% | 3.1% | 4.1% | 100% |
| *All the questions were voluntary, and eight participants chose not to provide self-evaluation |

39% of the participants chose the middle option (3) making it the most prevalent selection by a notable margin. We can consider two possible explanations for this outcome, which are not mutually exclusive. First, while many the students found it difficult to explain what AI is or how it works (questionnaire items 1 and 3), they were able to provide examples of where and why AI is used in society (questionnaire items 2 and 4). For instance, Student 83, whose self-evaluation was 3, responded "I don't know" to items 1 and 3 but mentioned that AI is used in "machines" (item 2: describe where you think artificial intelligence is or what it is used for.) in order to "receive knowledge" (item 4: describe why you think artificial intelligence is used). Student 83 also provided a couple of examples ("phone", "computer") to item 5 (name any words, things, or objects that you think are related to artificial intelligence), which clarified the kinds of machines the student was referring to in the previous answer. Thus, metaphorically speaking, choosing the middle option represents the "average" of the student's responses to items 1–5 (low knowledge about what AI is and how it works combined with higher knowledge about the practical applications and contexts of AI).

The second explanation is more methodological. A recent review (Coombes et al., 2021) suggests that children and adolescents sometimes have difficulties using the middle scales in surveys. In our case, it is possible that the students have approached the middle option in two different ways, resulting in a relatively high number of students choosing it as their option. Some students may have interpreted it to stand for moderate knowledge (as suggested in the example of Student 83, above), while others may have treated it as equivalent to "I don't know" or "no opinion" (see Wetzelhütter, 2020, for a more in-depth discussion).

### 4.1. Non-technological AI

The prefix "non-technological" refers to the kind of misconceptions in which AI was conceptualized as something else entirely than technology. Despite the semantical variation in students' responses, all the non-technological misconceptions can be understood as factual misconceptions: failed attempts to reason about the essence of AI. Student 84*, for instance, wrote that AI means that "one remembers things" with no references to technologies in any of his/her answers. As the excerpt illustrates, AI was conceptualized as a kind of cognitive process or an act or an action that people engage in.

Some students equated AI with the practice of forethought or premeditation. Student 50 wrote that "Artificial intelligence works so that you think about what you could do before doing anything". A similar view, if a bit more implicit, was present in Student 52's answer: "Artificial intelligence, in my view, means that one knows and understands what one is doing". Put differently, here the students connoted the term AI with actions in which people regulate their immediate and intuitive instincts and, instead, act based on a careful reflection of the situation. The students' lines of reasoning remind of Daniel Kahneman's (2011) famous distinction between two different modes of human thinking; "system 1" and "system 2". System 1 "operates automatically and quickly, with little or no effort and no sense of voluntary control", and —thus— it is the way of thinking that comes from us by nature (Kahneman, 2011, p. 20). The use of the system 2, in turn, requires "effortful mental activities", such as concentration (2011, p. 22). In other words, Student 50 seemed to think that processes where people resist their intuitive thinking are "artificial", because it requires us to resist our "natural" way of thinking. Other students had somewhat contrasting views to the above as they connoted AI with rapid and intuitive decision-making instead. Student 177 expressed that AI means that "Human comes up with a quick knowledge [like an answer to a question] in a difficult situation". Student 184 expressed a similar stance by commenting that AI could be about making "quick decisions".

One group of students stated that AI is about people's conscious and sometimes intentional act of faking or pretending that they are cognizant about things of which they actually know from little to nothing. Put differently, for such individuals, intelligence is artificial if the person in concern actually does not know what s/he claims to know — a misconception neatly captured in the following excerpt from "Artificial intelligence is used in things where you have to say something quickly even if you don't know it. artificial intelligence means self-invented matter" (Student 156). Student 158, in turn, expressed a bit different view as s/he stated that "AI means that you think you know something, but you actually don't. You sort of think that you are intelligent". In other words, in the student's view, intelligence is artificial if people think they are cognizant about things of which they are actually ignorant —a description that reminds of the Dunning-Kruger effect.

### 4.2. Anthropomorphic AI

A common misconception was conceptualizing AI as an anthropomorphic technology. In this category, students attributed AI with human-like feelings, mental states, or behavioral characteristics. Student 142*, for instance, stated that "AI works in such a way that robots can think" – a view shared by Student 143, according to whom AI means that "some device [has] similar intelligence and knowledge as humans". Such views can be categorized as vernacular misconceptions, because the students make no difference between AI and human intelligence. Anthropomorphism was not restricted to intelligence and knowledge only, but students also associated AI with other human-like characteristics. One student suggested that "AI can, for example, have its own emotions and personalities" (Student 21) whereas other students referred to AI explicitly as a "lifeform" (Student 24*) or a "species" (Student 62).

On one hand, such views may tell about non-scientific misconception that results from information retrieved from non-scientific sources: some students referred AI-imaginaries of popular culture either in explicit ("Jarvis, the artificial intelligence made by Marvel Studios for the movies" [Student 196*] or implicit manner ("one day, it [AI] will destroy the world", Student 123). On the other hand, anthropomorphism can also signal a conceptual misconception that draws from limited (conscious) experiences of and narrow focus on AI. Children often name voice assistants like Apple's Siri and Amazon's Alexa as an AI application they are familiar with (Kim et al., 2023; Kreisens & Schultz, 2021; Mertala et al., 2022) – a theme that was echoed in our data as well: "There are many types of AIs, for example Apple phones have an AI called Siri" (Student 24*).

Voice assistants frequently use expressions, which mimic that they would possess personality, emotions, and values. If the user asks whether Siri is a robot, Siri may reply that "I'm not sure what you have heard but virtual assistants have feelings too" (Taubenfeld, 2023). The purpose of such witty answers, of course, is not to obfuscate the user to believe that Siri is a sentient subject but to make the user experience fluid and enjoyable. However, there is emerging evidence that "humanizing" AI may steer children to add anthropomorphic attributes to the functionalities of AI-based applications, namely voice assistants. Some children in Szczuka et al (2022), for example, believed that Amazon's voice assistant Alexa is taught in a similar manner as humans, which implies that Alexa would learn, know, and understand things like humans do (which implies a presence of a vernacular misconception as well).

### *4.3. AI as a machine with pre-installed knowledge or intelligence*

The third form of misconception identified was AI as a machine with pre-installed knowledge or intelligence. In this category, AI was not seen to be able to learn literally or metaphorically. Instead, the information AI processes —and provides for the user— is saved or installed in the machine or in the software in advance. Student 10*, for instance, wrote that "AI means knowledge that are feeded into the computer". Student 188 expressed a similar idea by stating that "In my opinion AI is pre-installed knowledge in robots, for example. AI is not learnt knowledge" — a view shared by Student 55* according to whom "AI is a human-invented intelligence that is often given to robots". Similar views were expressed about AI's intelligence as well, as illustrated in the following excerpts: "Artificial intelligence is made intelligence, not intelligence that has manifested by itself" (Student 49); "I think artificial intelligence means that it is, for example, a robot that has been made intelligent" (Student 162).

It is certainly true that AI-based computational models do not learn in a similar manner as people do. However, many AI applications are capable of improving their accuracy and efficiency through supervised or unsupervised machine learning when new (and often better) data is entered into the system. This means that the applications' capabilities are not static but subject to evolving over time –a process, which in the context of humans would be described as learning or development.

Because the students conceptualized AI as a technology whose information processing differs from humans without understanding the functional principles of AI, this category was interpreted to be mainly about factual misconception. That being said, the Finnish language might play a role in the emergence of this particular category, which implies the presence of vernacular misconception as well. In Finnish, the computer is called "tietokone", a verbatim translation of which would be a "knowledge machine". In other words, unlike in English, there is no linguistic reference that a computer would perform "computation" in Finnish; the closest translation for computing ("laskenta") connotes more to "calculation", and another option ("tietojenkäsittely") translates to "information processing". Instead, in the Finnish language, there is an in-built explicit reference that a computer would "know" things. Indeed, juxtapositions between AI and computers as "knowledgeable machines" were present in the data: student 135 provided an explicit reference by writing that "I think artificial intelligence means a device that doesn't think but knows things. Just like a computer does". We will discuss the role of the linguistic cues in more detail in the concluding section.

### 5. Concluding remarks

In this study we have explored Finnish 5th and 6th graders misconceptions of AI. Three categories were formed via an abductive analysis of qualitative open-ended survey data: 1) non-technological AI, 2) anthropomorphic AI, and 3) AI as machines with preinstalled knowledge or intelligence. Our findings carry similarities with previous related research. Especially anthropomorphic misconceptions are identified also in previous AI- and computer-themed conception research (e.g., Kim et al., 2023; Kreinsen & Schultz, 2021; Rucker & Pinkwart, 2016). The same applies also to understanding AI and computers as databases with pre-installed knowledge (Kim et al., 2023; Rucker & Pinkwart, 2016). Likewise, the number of students who evaluated their AI-knowledge as high (18.4%) is roughly similar with Selwyn's et al (2022) findings from Australian adult population (25.5%) (especially considering that unlike in Selwyn et al., 2022, none of our participants naturally possessed degree in computer science). Furthermore, similarly to Kim et al. (2023) various misconception types (factual, conceptual, vernacular, and non-scientific misconception) were present in our data.

### 5.1. The emergence of non-technological misconceptions

The non-technological misconceptions, in turn, are a novel —and even surprising— finding. They are surprising in the sense that AI is a common theme and concept in news media (Slotte Dufva & Mertala, 2021) and popular culture (Cave & Dihal, 2019). Thus, our initial assumption was that the students would have heard the term and connotated it to digital technology. Indeed, many of the students who were not able to explain what AI is or how it works were still able to provide concrete examples of its applications by referring to smart phones, computers, games, social media, robots, and others when asked where and why AI is used. That, however, was not always the case, which is neatly summarized in Student 177's comment "I have not heard this word [AI] before. Quick decisions?".

There are, at least, three different yet not mutually exclusive explanations for the emergence of the non-technological misconceptions. The first one is that our sample (N=195) was notably larger than that of 8 to 17 in previous research on children's conceptions of AI (Kim et al., 2023; Kreinsen & Schultz, 2021; Ottenbreit-Leftwich et al., 2021, 2022; Solyst et al., 2023; Vandenbergen & Mott, 2023). The larger sample, arguably, increases the diversity in participants' views and experiences.

The second explanation is that we aimed to avoid a priming effect in the instructions (see, Selwyn et al., 2022) as the way how the concept is introduced and framed to children arguably affects their responses. Put differently, it is difficult to imagine that students of similar age would possess non-technology-related misconceptions about machine learning since the concept includes the actual word "machine". On the other hand, the term "machine learning" in itself is not explicit about who is learning and from whom whereas an instruction to explain "how one could teach a computer" (Vartiainen et al., 2021) provides children explicit cues about the direction of the relationship. In other words, with less nudging instruction, machine learning could be also understood as a process in which a human uses a machine, a computer, for instance, for learning purposes.

The third explanation relates to language. In the Finnish language, AI is called "tekoäly", a compound word that unites the terms "teko" and äly". While the latter term, "äly" translates as intelligence, "teko", is a more ambiguous concept: besides "artificial", the word also refers to "an act" and "an action". Indeed, many of the non-technological misconceptions described different kinds of cognitive acts and actions (e.g., regulation of immediate and intuitive instincts). This explanation is supported by Mertala's (2019) finding that Finnish preschoolers used conceptual similarities as the basis of their reasoning of what programming is, as in Finnish the words for programming ("ohjelmointi"), program ("ohjelma"), and manual ("ohje") are notably similar.

### 5.2. Limitations and suggestions for future research

While our findings provided novel and useful information about children's misconceptions about AI, the study is not without its limitations. The obvious limitation of the use of survey method is that it prevented us from asking clarifying questions from the participants. It should also be noted that our sample is not representative, and therefore, the findings, especially the distribution of different misconception categories and self-evaluated AI knowledge, cannot be generalized for the whole population. That said, the current stage of qualitative research on students' AI conceptions (e.g., Kim et al., 2023; Kreinsen & Schultz, 2021; Mertala et al., 2022; Ottenbreit-Leftwich et al., 2021, 2022; Solyst et al., 2023; Vandenbergen & Mott, 2023; Vartiainen et al., 2021) seems rich enough to provide well-justified theoretical foundations for large(r)-scale quantitative and/or mixed-method studies, which we encourage future research to carry out.

Additionally, when making conclusions based on our findings, it is important to note that placing the focus on misconceptions tells only a partial tale about the variety of conceptions the students had. In fact, one student argued that AI is "replication of human activities with machines that don't fill the criteria [of being a human]" —a definition, which contains notable resemblance with Kurzweil's (1990) idea. Additionally, many students expressed quite elaborated knowledge about the functional principles of different kinds of AI-solutions. They, for instance, noted that AI-applications often use different kinds of sensors to collect data. These findings are reported in another publication (Anonymized).

Another limitation worth considering is that the data was collected from one country only. Previous research has identified variations in attitudes toward AI and autonomous robots by people living in different geographical and cultural areas (e.g., Dang & Liu, 2021; Druga et al., 2019). Comparative studies are needed to gain a more holistic understanding of students' (mis)conceptions of AI as well as to gain more comprehensive knowledge about the role of different languages in children's AI mis- and preconceptions. As discussed in the Section 5.1., it is possible that due to the specific linguistic features, non-technological misconceptions (at least in the forms presented in this study) may be restricted to certain geographical and cultural areas, namely Finland. However, the concept of machine learning does not bear similar context-bound lexical complexities and, thus, provides an interesting case for cross-cultural comparative research.

It should be also noted that the data fails to tell much about the foundations of the misconceptions. While our findings imply that conceptual connotations, media representations, and hands-on experiences with AI solutions like voice assistants have a role in students' misconceptions —an interpretation supported by previous research (e.g., Cave et al., 2020; Mertala, 2019; Mertala et al., 2022)— more research is needed to verify (or confront) these observations. Lastly, at the time of the data collection (Spring 2021) the everyday AI-landscape was somewhat different than it is today due to the widespread and rapid proliferation of generative and conversational AI-solutions like ChatGPT. Additional research is required to explore whether the recent developments and the rather sensational media discourses around ChatGPT, Google Bard and others have shaped children's pre-instructional conceptions of AI.

### 5.3. Pedagogical implications

Despite these limitations, our findings provide implications for AI-literacy education. It may be helpful to outright lexically deconstruct terms like computer ("tietokone"), AI ("tekoäly"), and machine learning ("koneoppiminen") and discuss how the words themselves may be conceptually misleading. This notion is not restricted to Finnish only. In Estonian, AI is "tehisintellekt", which verbatim translation "power intelligence", arguably, provides different connotations than its English or Finnish counterparts. That said, paying attention to the language around AI is important on a more universal scale as well.

Linguistic Emily Bender (2022 November) has suggested a thought experiment in which the word "AI" is replaced by "mathy maths" in situations where cognition-related verbs like "think", "decide" or "understand" are used to describe AI's functions. While her initial idea was to help people to unpack the (commercial) hype talk around and about AI, we believe that a similar play with words could also help students to critically reflect on their initial anthropomorphic misconceptions (while simultaneously supporting their critical media literacy as Bender intended). Indeed, clearly informing students about the fictitiousness behind the conceptions of non-technological AI or anthropomorphic AI should not be held redundant but perhaps even as core elements in introductory AI literacy education. Especially becoming aware of media influences —from which Bender draws her examples as well— seems a meaningful starting point (see also Mertala et al., 2022).

Bender's critique also raises questions about the role of the concept "AI" in AI literacy education, and the term is not universally embraced (e.g., Jordan, 2019; Lainer, 2023 April; Mason, 2021 January). According to Jaron Lainer (2023 April), the concept "AI" is misleading (a notion supported by the findings of this study). To cite Lainer's (2023 April) actual words: "It's easy to attribute intelligence to the new systems; they have a flexibility and unpredictability that we don't usually associate with computer technology. But this flexibility arises from simple mathematics" (or "mathy maths" to use Bender's terms). Bottom-up approaches in which the students are gradually provided an overall picture of AI (as a field) by exploring its practical applications like object recognition, selection trees, and others, and by using an explaining the concepts systematically, could be one practical way to demystify and concretize AI as part of AI literacy education.

Indeed, due to the widespread and rapid proliferation of generative and conversational AI-solutions in various areas of everyday life (including education) it is vital to acquire a more in-depth understanding regarding how AI systems actually work, for instance, by becoming familiar with concrete methods of machine learning (and teaching a machine) (see Vartiainen et al., 2021). This should help students to understand how AI requires data and that it operates solely based on said data by (often) identifying patterns and calculating probabilities. In other words, while capable mimicking a human-like two-way interaction, conversational AI does not think similarly to people (anthropomorphic AI) nor does the device have preinstalled knowledge or intelligence (AI as a machine with pre-installed knowledge or intelligence). By making these differences visible, teachers can provide a glimpse behind the curtain of highly abstracted everyday AI user interfaces (e.g., seemingly sentient and complex emotional personalities, like Siri). Furthermore, the questions about AI ethics – a central theme in various AI literacy curricula (e.g., Kim et al., 2021; Payne, 2019) are fundamentally different whether a student thinks that solutions made by an autonomous AI-technology —like a self-driving car— are based on mathematical models (and the ethical solutions behind them) or subjective emotions of an anthropomorphic machine.

The current stage of research suggests that there can be a great variety of mis- and preconceptions among students when taking their inaugural steps in learning about AI (e.g., Kim et al., 2023; Kreinsen %Schultz, 2021; Mertala et al., 2022; Solyst et al., 2023; Vandenbergen & Mott, 2023). In other words, educators should be mindful of a potentially large gap between students' nascent ideas and the scientific nature of how and why AI works and what it is. That being said, it is important to acknowledge that the vast majority of the students were rather skeptical regarding the depth of their AI-knowledge as only 18.4% of the participants expressed that they are well aware about AI. For example, many students with non-technological misconceptions evaluated their knowledge about AI either low (1–2/5) or moderate (3/5). This implies that while the misconception is quite fundamental, it should not be that difficult to fix since the degree of certitude (Usó-Doménech & Nescolarde-Selva, 2016) is low. Indeed, there is evidence (see, Kim et al., 2023; Mertala, 2020) that superficial technology-related misconceptions can be replaced with (more) correct ones via rather simple pedagogical practices.